# Recent Experimental and Theoretical Advances in Microdrilling of Polymers with Ultraviolet Laser Beams

## Sylvain LAZARE and Vladimir TOKAREV

*LPCM UMR 5803, Université de Bordeaux 1, 351 cours de la Libération, F-33405 Talence, FRANCE*
*E-mail:  s.lazare@lpcm.u-bordeaux1.fr*

## ABSTRACT

Laser drilling becomes of increasing importance when hole diameter is in the range of 10 to 50 µm, for which conventional alternative approaches are becoming difficult and cost inefficient. Furthermore, it is viewed as the technique of choice for a number of composite materials and hard materials which are not readily processed at the microscopic level by contact mechanical tools. We have demonstrated that suitable experimental conditions are capable of producing microholes with record aspect-ratio (up to 600) in pure polymers like PET, PI, PC, PS, PMMA, PEEK, .. . For example holes of diameter typically 30 µm can be as long as 18 mm, depth at which the drilling rate is getting nearly zero and the profile stationary. Other materials (metals, ceramics) which can be similarly laser microdrilled do not exhibit such very high aspect-ratio. The mechanisms of the drilling process have been studied in details and an original analytical model has been constructed recently. The various experimental results, obtained with the KrF laser, will be reviewed with emphasis on the parameters leading to formation of good holes with high aspect-ratio. For the application it is also important to note that such high values of aspect-ratio are obtained with regular configuration of the KrF laser giving a standard divergence of 3 mrad. However as shown by the model there is still room for improvement by using a beam with a lower divergence (theoretical limit is 0.2 mrad). Further experimental work is now in progress.

**Keywords:** Materials, laser, processing, drilling, model, profile, mechanisms, polymer, divergence, cutting

## 1. INTRODUCTION

Microdrilling is one of the key processes for future technology and industry. Conventional mechanical drilling has to face increasing physical limitation when hole diameter is decreased. Below some critical dimension, friction surpasses the mechanical strength of the tool and machining is possible only by choosing specially designed tools made of high modulus materials (e.g. diamond, cBN, etc..) Additionally processing time may be increased and tool lifetime may be sacrificially reduced. In any case it means that the cost of the process can be prohibitively increased. In this context laser microdrilling [1-11] offers an attracting alternative to mechanical machining and is already widely used in numerous applications. In particular the same laser tool can process similarly any material regardless its physical properties (hardness, softness, composite, absorptivity .. ), although mechanisms of material removal can be very diverse. In this framework we have recently [12-16] investigated the ultimate limits of laser microdrilling mainly in polymeric materials. In particular with respect to the two dimensions of interest in microdrilling applications, hole diameter (*d*) and maximum hole length (*l*) are obtained for a given set of laser parameters (wavelength, beam size, energy density). In the present work we show that unprecedented high aspect ratio (*R*=*l*/*d*, up to 600) can be reached when experimental conditions are adequately chosen. We have experimented mainly with polymers however results can be transposed to other materials (metal ceramics..) with more or less complications. The first consideration can be guided by the simple equation (1) giving laser waist size *d* after transformation of a gaussian laser beam of initial diameter *D* by a single lens (**Fig.1**) of focal length *f*  [17]:

$$d = 4\lambda f / \pi D \qquad (1) \qquad (\lambda \text{ is the laser wavelength}).$$

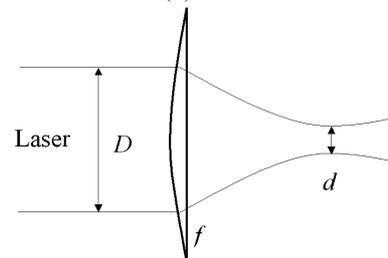

**Fig. 1** Beam transformation by a single lens

Small beam diameter is favoured by choosing a short laser wavelength and small focal length. To illustrate this: a beam of 3 µm is obtained with a KrF laser ($\lambda$=248 nm) with a lens of 20 mm focal distance for an incident beam size of 1 mm. For comparison an infrared $CO_2$ laser beam ($\lambda$=10.6 µm $\approx$ 43x0.248 µm) would give a 43 times larger spot ($d$=120 µm). For the same initial laser pulse energy the fluence at the waist would be accordingly much lower ($\sim(\lambda_{KrF}/\lambda_{CO2})^2 \approx 1/1850$) because spread over a larger surface area. Therefore ultraviolet lasers are better candidates for making small spots of high energy density in order to perform high aspect ratio holes by deep ablation. Moreover in UV, thresholds of ablation are low because absorption coefficients $\gamma$ are high ($\sim 10^5$ cm$^{-1}$). Therefore energy density thresholds $F_t$ (threshold fluence) can be easily surpassed and the large excess of energy with respect to the threshold value provides the large kinetic energy necessary to expulse the ablated material outside the deep hole. In the present state of the art, powerful KrF lasers (0.4 J/pulse, 200 Hz) are mainly used because of the high demand in pulse number of microdrilling, but other shorter wavelengths (e.g. ArF, or VUV) are also considered although suffering of lower energy per pulse and/or lower repetition rate. It is also envisageable that, in the future, laser with even shorter wavelength (far-UV, X-ray) will be considered owing to their large potential of focusability. Other advantage of ultraviolet radiation is the high energy content of the photon (e.g. $h\nu$ = 5 eV for 248 nm) which is large enough to cause direct bond breaking by single photon absorption. Polymers are long chain molecules. Part of the incident energy is converted into heat giving rise to phase transition like solid-to-liquid and complex mechanisms of liquid expulsion can also take place. Large wavelengths lasers, having photon energy smaller than average polymer bond energy, gives higher tendency to thermal mechanism, less direct decomposition into gas and larger affected volume around the keyhole as explained below. Therefore although efficient drilling is obtained with such lasers in a wide range of applications, the best performances in terms of microscopic precision (lateral diameter) are now obtained with short wavelength lasers. The efficiency of ultraviolet laser of a given energy having nanosecond pulses in microdrilling is then due to the synergy of high focusability and high absorption coefficient usually encountered in solid materials which gives low ablation threshold and making possible ejection of material with the high kinetic energy to deep drilling. This paper starts with a description of the experimental setup stressing the strategic points. From the experimental results reported in part 2 below we will see that maximum aspect ratio increases with fluence as predicted by the model presented in part 3. The model also predicts hole profile, beam transmission in the keyhole and a polymer fluence threshold which could well be the minimum energy to significantly remove some materials after incubation or ablation with many pulses. In following part 4, mechanisms and perspectives will be discussed.

## 2. MICRODRILLING EXPERIMENT

In these experiments good microdrilling is obtained only when the laser microbeam has a high quality over the large depth of the wanted hole, i.e. high energy density, smooth profile and small lateral dimensions. Microdrilling reveals to be in itself a good test of the beam quality that can easily be used by the experimentalist for adjusting the optical setup. Any other photon detector (like a camera) is in the present case useless or inconvenient since the density of energy used is too high and the size of the beam too small.

### 2.1 Laser beam setup

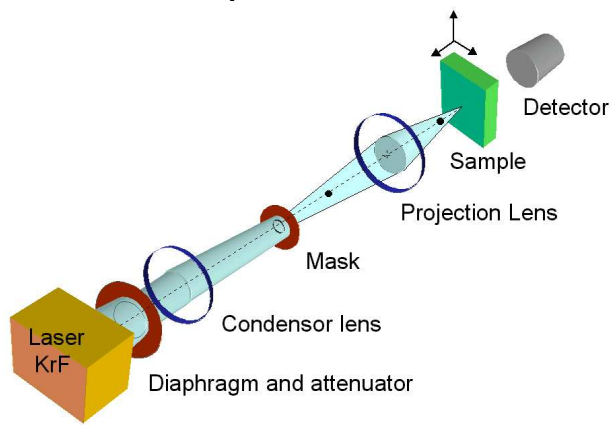

**Fig.2** Optical setup used for the microdrilling experiments

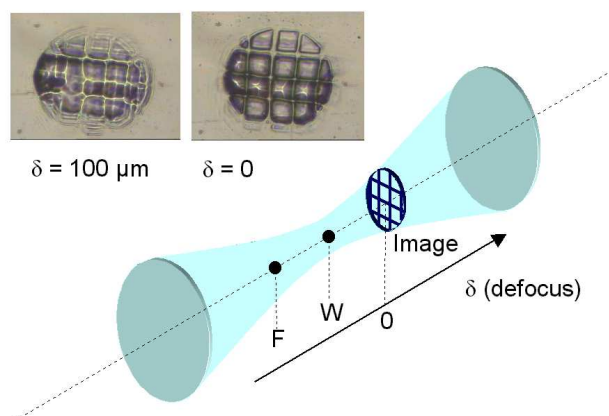

**Fig.3** Positioning of the sample in the beam waist with the aid of a nickel grid placed at the mask whose image is looked for in the waist by ablation with one pulse on PC surface. $\delta = 0$ is by definition the image position.

The laser beam setup is shown in Fig.2. It is composed of two silica (Suprasil) plano-convex lenses. One is the condensor ($f_{cond}$=250 mm) and the other is the projector (f=25 mm). The are chosen of small diameter in order to keep the thickness at the beam path to a minimum and avoid unwanted attenuation of the radiation (they must stand 200 pulses/sec for minutes with good stability). The use of a precision lens [18] as projection optics is not completely excluded, but was not preferred for this particular study. The excimer laser is a Lambda Physik LPX 220i (pulse energy $E$=0.4 J, pulse duration $\tau$=25 ns, pulse peak intensity $I$=$E/\tau$=1.6x10$^7$ W and repetition rate 200 Hz maximum) whose cavity is equipped with regular planar aluminium mirrors and which has a length of about 1.50 m, therefore offering a maximum of energy in the low divergence resonator modes. A circular molybdenum aperture (e.g. $\Phi$ = 250 μm) is used to shape the beam at a distance of the condensor equal approx. to $f_{cond}$. This mask is precisely imaged onto the polymer surface, as illustrated in Fig.3, with the projection lens so that the demagnification ratio is of the order of 1/5. A fine nickel grid (Goodfellow, period 340 μm) is placed on the mask and the polymer target is moved near the waist position up to obtaining a good position giving the best ablation pattern (which is then called defocus zero $\delta$= 0 )as seen in the optical microscopy image in upper left inset in Fig.3. Polycarbonate is used for this fine adjustment made with one single pulse, since its ablation is clearly not disturbed by liquid formation and lateral flow along the surface. It proved to be a very sensitive technique. Two diaphragms are also used to limit the influence of scattered radiation, which are positioned before each lens. Their adjustment is done on an empirical basis.

## 2.2 Polymer samples

The selected polymer samples were chosen in a first series because of their good optical and surface quality (PC, PS, PMMA, PET) allowing an easy optical microscope observation and in a second series for their technological importance (PI and PEEK). The polymers studied are polyimide (PI), poly(ethylene terephthalate) (PET), polystyrene (PS), poly(ether ether ketone) (PEEK), poly(methyl methacrylate) (PMMA) and bisphenol A polycarbonate (PC). Commercial names are Kapton for PI, Mylar D for PET, Stabar for PEEK. PS samples were taken from Petri dish and PC samples from polycarbonate panels Axxis (DSM Engineering Products). As seen from Table 1, PI, PET and PEEK have high absorption coefficient, whereas PMMA, PS and PC are weak absorbers at room temperature for the used laser wavelength. It gives the possibility to study the effect of this important parameter on the results of drilling.

**Table 1**  Polymer absorption coefficients $\gamma$ at the KrF laser wavelength 248 nm, skin depth 1/$\gamma$, ablation threshold $F_t$ at 248 nm and drilling parameters: rate $D_r$ maximum aspect ratio $R_m$ and conicity of the hole α.

| Material | $\gamma$(μm⁻¹) | $1/\gamma$(μm) | $F_t$ (mJ/cm²) | $D_r$ (μm/pulse) | $R_m$=d/l | $\alpha$=1/$R_m$ (10⁻³) |
|---|---|---|---|---|---|---|
| PET | 16 | 0.065 | 30 | 0.7 | 565 | 1.77 |
| PC | 1 | 1.0 | 40 | 0.8 | 390 | 2.56 |
| PEEK | ~10 | ~0.1 | 50 | 0.6 | 385 | 2.60 |
| PI | 22 | 0.045 | 54 | 0.4 | 360 | 2.78 |
| PS | 0.61 | 1.6 | 40 | - | 315 | 3.17 |
| PMMA | 0.0063 | 150 | 250 | 2.5 | 255 | 3.92 |

## 2.3 Rate of drilling and stationary profile with PET example

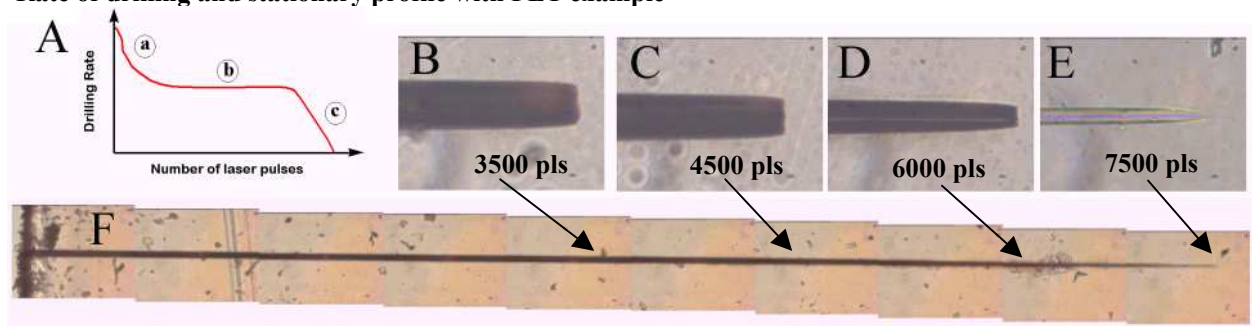

**Fig.4**  Microdrilling characteristics of a hole with 25 μm diameter and 5 mm length drilled in **PET** with the KrF laser. **A** Drilling rate of advancement as a function of time (or pulse number) **B** to **E** shape of the advancing ablation front at various times of the experiment. **E** and **F** hole after stationary profile (end of microdrilling), **E** Final sharp tip. The scale of the pictures is given by the hole diameter 25 μm and the hole image results of the assembling of many partial images of the same hole.

A typical microdrilling experiment, done after beam optimisation, is illustrated in Fig.4. It requires usually many pulses. As in Fig.4 **A**, the rate has an initial value (a) which can be related to surface ablation with 3D expansion characteristic of surface ablation (as opposed to deep ablation), then it displays a constant value (b) in deep keyhole because plume absorption is increased and redeposition sets in, and finally it goes to a "zero value" (c) when the stationary profile is reached. As in Table 1 with an average rate of 0.7 µm/pulse the stationary profile is reached in approximately 7500 pulses. The intermediate pictures **A,B,C** are respectively taken at 3000, 4500, 6000 pulses. We observe the evolution of the main ablation front or hole bottom which is rather square at the beginning and which becomes sharper when the sta-

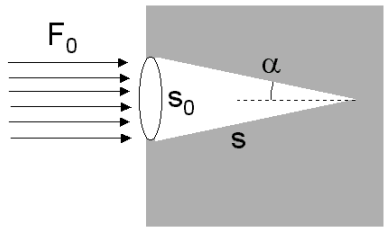

**Fig.5** Simplified model of stationary state of a drilled conical hole.

tionary state or end of drilling is approached. When the stationary state is reached the tip does not move anymore, the lateral dimensions of the hole do not change as well. The laser energy is therefore trapped within the hole and spread over the hole inner surface so that its surface density is lowered below ablation threshold. This result can be better understood with a simplified model of a conical hole in Fig.5 (we will see below that the hole profile is not conical). An ideal fully collimate beam however would create the formation of a cone of angle α and aspect ratio 1/α. The drilling then would stop when the fluence at the wall is a certain threshold $F_\infty$ so that $F_0 s_0 = F_\infty s = F_\infty s_0 / \alpha$ (2). From this aspect ratio can be estimated for example as follows: $R = 1/\alpha = F_0/F_\infty > F_0/F_t$

(3), since the threshold $F_\infty$ at which the drilling stops is lower than the usual surface ablation threshold $F_t$. With typical fluence $F_0 = 30$ J/cm² and threshold $F_t$ = 30 mJ/cm² , an aspect ratio of the order of $10^3$ would be predicted. This is comparable to what experience gives.

### 2.4 Other polymers

PET under the form of Mylar D, presented above gives the best aspect ratio (Table.1) then there is a group with PC PEEK and PI which displays lower aspect ratio $R_m$ ~350 and PMMA, PS have the lower values. It is interesting to understand why, when chemical structure is varied, the microdrilling performance is so strongly affected. This particular point will be discussed below with the presentation of drilling features related to each specific polymer.

#### 2.4.1 Polycarbonate (PC)

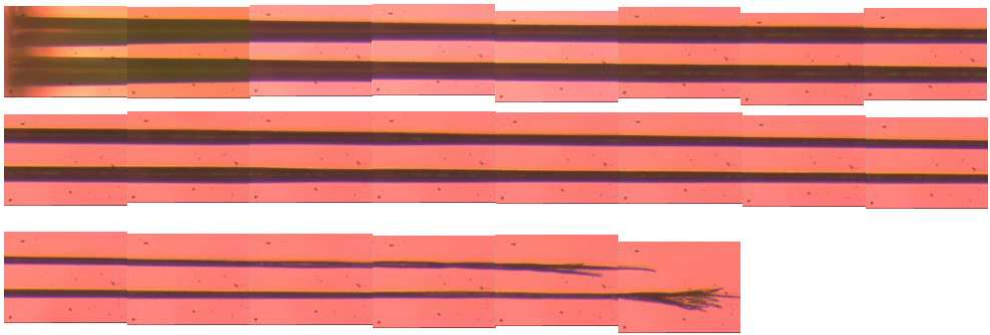

Polycarbonate also gives high aspect ratio holes (Table 1). The surface after ablation has a black color characteristic of carbon material, already observed in multipulse surface ablation of PC. As we already reported at some point of the drilling a branching phenomenon can be observed. In Fig.6 it appears at the tip end when the laser fluence is attenuated to values in the vicinity of the threshold value. The mechanism of formation of the branches [16] is analogous to the formation of cones, reported in surface ablation. They form because the absorbed fluence $FA(i)cos(i)$ becomes equal to $F_t$ ($i$ is the beam incidence angle). This branching is a specific phenomenon of polycarbonate and other polymers like PI and PEEK. They hold in common the characteristic of yielding by ablation graphite to a large extend, whereas other polymers PET, PS and PMMA much less. It is also thought that expulsed ablation products are gases and nanoparticles as opposed to liquid micro and nano droplets in other cases.

**Fig.6** A set of two adjacent holes in polycarbonate of 25 µm diameter and distant of 40 µm showing the precision of the laser drilling. The second hole drilling is not perturbed by the presence of the first one.

### 2.4.2 Poly(methylmetacrylate) (PMMA)

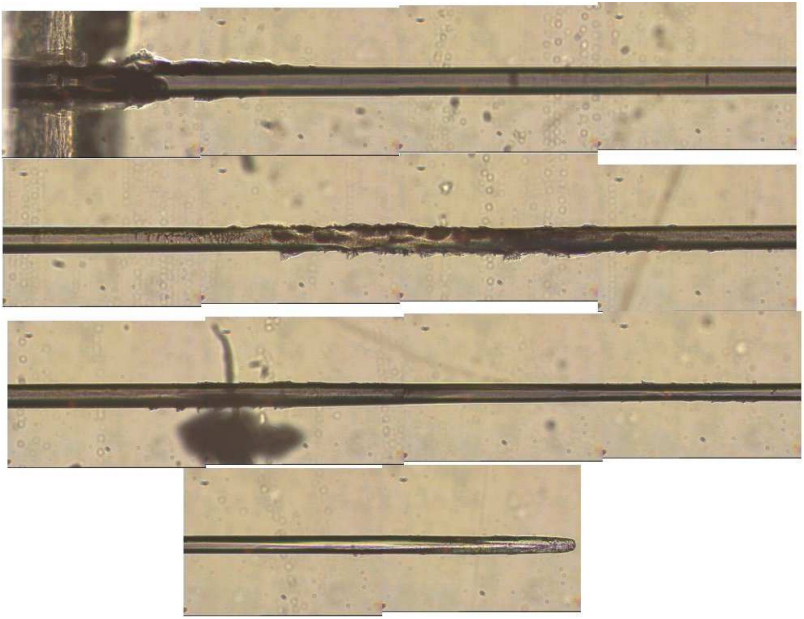

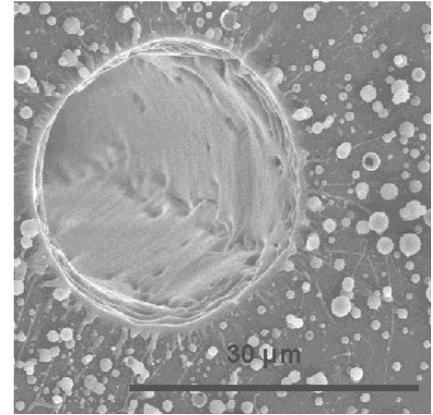

**Fig.8** SEM picture of PMMA drilled with the KrF laser. The hole entrance diameter 30 μm. It shows many redeposited polymer droplets originating from the liquid expulsion.

**Fig.7** KrF laser drilled hole in PMMA viewed with optical microscope. Diameter 60 μm and length 6.5 mm (AR=100). Top left is entrance and bottom right is end tip. The total image is assembled from several partial pictures.

An example of KrF laser drilled hole in PMMA is shown above in Fig.7 with a magnification which allows to reveal several interesting features concerning the drilling mechanism. Its aspect ratio is only $R$=100 in the Fig.7 and not the maximum aspect ratio $R_m$=250 (Table 1) because its diameter does not allow to reach the fluence value high enough to reach $R_m$. A smaller diameter beam (~30-35μm) obtained with a smaller molybdenum mask (30 x 5 = 150 μm) yields the more energetic spot capable of reaching the maximum aspect ratio $R_m$ of Table.1. One main difference with the previous case of PC is the shape of the end tip which is now completely round instead of branched. This is due to the presence of liquid although viscous PMMA polymer during the laser interaction. The liquid surface tension produces a meniscus at the bottom of the hole which resolidifies between two consecutive pulses. Such liquid is not produced in drilling of the other polymers PC, PI, PEEK and since it is formed an ultrathin layer [19,20] in the ablation of PET, it seems to have a negligible effect on its drilling. SEM imaging in Fig.8 displays the redeposited PMMA droplets around the hole entrance. Liquid expulsion may be a significant mechanism of material removal in the case of PMMA. Recently we also demonstrated the experimental conditions in which nanofilaments [21,22] are formed by droplet expulsion with a single laser pulse. In these cases speed of liquid ejection can be as high as 840 m/s (~Mach 3) according to our recoil pressure model and depending on the fluence used. Details of the hole surface can be seen in Fig.7 where in particular some microcracks are visible. It is difficult at this point to explain their mechanisms of formation but their presence is not surprising since the PMMA material is exposed to a large dose of UV and the P ant T pulses in the hole during microdrilling have large amplitude and impose some strain to the target. Also due to the large penetration depth ~150 μm involved in PMMA (Table.1) we can imagine that a cylindrical shape affected volume is formed around the hole where molecules receive a lot of photonic and thermal excitation. This volume was measured by microraman spectroscopy.

### 2.5 Search for the optimum aspect ratio

As included in Table.1, there exists a maximum aspect ratio $R_m$. It is mainly a function of polymer type when drilling is performed with the same beam for all polymers. But it is also sensitive to beam diameter since beam fluence is the main controlling parameter in these experiments. We determined that aspect ratio varies with the beam fluence like :

$R = R_m(1 - e^{-F/F_c})$ in which $F_c$ is a fluence characteristic of the polymer [12]. Now the fluence which can be obtained at the waist center is the very important parameter since it should be as large as possible to drill deeper, as suggested by equation (3). It is made larger by decreasing the size of the beam but at some point the diffraction tends to spread the beam over a larger solid angle. This is the reason of the maximum aspect ratio $R_m$ that we can demonstrate in these experiments. It is obtained with a beam diameter of the order of 25-30 μm depending on the polymer and it is a limitation due to the laser beam characteristics.

## 3. MODEL

The analytical model presented below was described in details in previous publications [15,16]. The laser radiation approximated as in Fig.9a by a collimated beam such that it diverges from the focal point O of the projection lens P and it gives a spot of diameter $d$ at the surface of sample S positioned in the image plane of the mask M. So in the model, O is like a virtual point source (Fig.9b) shedding some light rays onto the sample and into the hole in a cone of half angle $\alpha_0$. The laser geometrical parameters are determined by the cone of light emitted by O in direction of the hole and defined by distance $z_0$, the cone angle $\alpha_0$ (numerical aperture of the beam) and the fluence profile (flat top, gaussian, etc..) which will be taken as a function of the inclination $\alpha$ of a particular ray (Fig.9b). We show by the present model that the hole geometry is predetermined by these fixed parameters. The model was validated by fitting the experimental hole length $l(F)$ with the predicted law. For some reason related to hole diameter increase at stationary profile the aspect ratio $R$ is not suited for this fit.

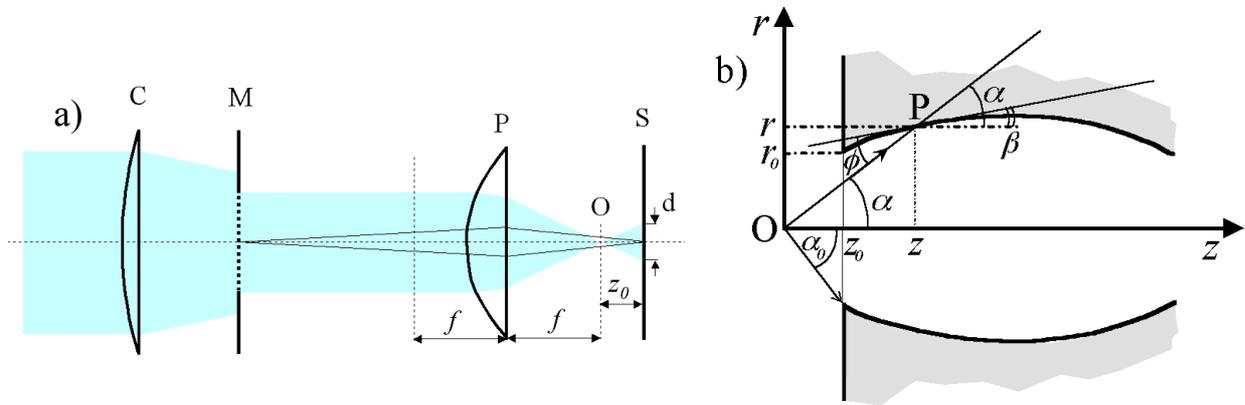

**Fig.9 a)** Schematic beam set-up used in the modelling of laser hole drilling. **b)** Point source model with ray propagation into the laser drilled hole. O is a virtual point source at distance $z_0$ of the sample and emitting rays in a cone of angle a0 into the hole with intensity varying with inclination a (laser profile). Hole diameter is strongly exaggerated for the sake of better viewing.

### 3.1 Condition of stationary profile

At the end of drilling the hole profile is stationary and therefore does not change significantly since ablation does not produce anymore. If it is so this means that the laser fluence absorbed at the hole surface $F_a$ has become smaller than the threshold value that we named $F_\infty$ constant over the entire surface. At point P of Fig.9b we therefore have the condition

$$F_a(r,z) = F_\infty \qquad (4),$$

the equation which characterises the stationary profile. $F_\infty$ is a threshold fluence at which the ablation stops, which depends mainly on the polymer and on its modification introduced by the laser dose during the drilling process. We will see that it is fairly different from the well known ablation threshold. It is simply a threshold below which the disappearance of the laser excited material does not produce anymore.

### 3.2 Hole profile at stationary stage

The absorbed fluence at point P with coordinates (r,z) of the profile (Fig.9b) is given by

$$F_a(r,z) = A_{eff}(\phi(r,z)) \times F(r,z) \sin \phi(r,z) \qquad (5)$$

where $F(r,z)$ is the incident fluence at P and $\Phi(r,z)$ is the grazing angle for ray OP. $A_{eff}(\phi(r,z))$ is a local effective ab-sorptivity for the laser radiation at point P. $F(r,z)$ is obtained from the angular fluence distribution $F(\alpha)$ at $z = z_0$ by

$$F(r,z) = F(\alpha)z_0^2/z^2 \qquad (6)$$

in which the factor $z_0^2/z^2$ describes the attenuation of fluence with distance $z$. The main two reasons responsible for ablation stop at stationary profile are then increasing attenuation with distance and increasing incidence angle with drilling progress. In equation (5) $\phi(r,z)$ can be expressed as a function of the ray inclination $\alpha$ (Fig.9b):

$$\phi(r,z) = \alpha(r,z) - \beta(r,z) \qquad (7)$$

where $\alpha = \arctan(r/z) \cong r/z$ since $r$ is small and

$$\beta = \arctan(dr/dz) \cong dr/dz = d(\alpha z)/dz = \alpha + z(d\alpha/dz) \qquad (8)$$ is the local inclination of the side wall at point P. Substituting (8) into (7) gives:

$$\sin\phi(r,z) \cong \phi(r,z) \cong -z(d\alpha/dz) \qquad (9).$$

Combining (5), (6) and (9) gives the differential equation determining the keyhole profile:

$$z\frac{dz}{d\alpha} = -A_{eff}(\alpha)\frac{z_0^2}{F_\infty}F(\alpha) \qquad (10)$$

Now in the following $A_{eff}$, the effective absorbtivity will be given the value 1 [16] on the basis of the following arguments. The laser radiation entered in the hole has very little chance to exit again back to the outside and multiple scattering occurs within the hole which makes the absorption of the radiation energy total. So at all point incident rays are finally absorbed regardless their origin since probably scattering dominates the beam propagation inside the hole. Furthermore attempt to fit experimental data with any other values than $A_{eff} = 1$ does not succeed to predict experimental measurements. With this value 1, the multiple scattering model equation (10) can be solved in the parametric form $r(\alpha)$, $z(\alpha)$ with $\alpha$ the parameter being the ray inclination: The equation of the hole profile at the final stationary regime is then given by:

$$r(\alpha) = \alpha z(\alpha), \quad z(\alpha) = z_0\left[1 + \frac{2}{F_\infty}\int_\alpha^{\alpha_\infty} F(\alpha')d\alpha'\right]^{1/2}, \qquad (11)$$

in which $\alpha_\infty$ is the limit angle above which integration is negligible. The hole length or depth can be infered from (11):

$$l(F) = z(\alpha = 0) - z_0 = z_0\left[\left[1 + \frac{2}{F_\infty}\int_0^{\alpha_\infty} F(\alpha')d\alpha'\right]^{1/2} - 1\right] \qquad (12)$$

### 3.3 Fit of the experimental hole length for top-hat beam profile

As stated above the hole aspect ratio was not taken as fitting quantity since it levels off at high fluence whereas the hole length continues to increase due to slow diameter increase with fluence which cannot be simply predicted by this model. If a constant diameter versus fluence is used besides the predicted length (12) then theoretical aspect ratio is an increasing function of fluence and does not reproduce the levelling off of the experimental data. On the contrary it was found that the hole length increases with fluence as predicted by equation (12) as seen on Fig.10. The experimental data were fitted successfully with the model (12) for the six polymers studied in this work. The top-hat beam profile is defined by:

$$F(\alpha) = const = F \qquad (13), \qquad \text{for } 0 < \alpha < \alpha_0 \text{ (see Fig.9b)}$$ and the hole length versus fluence dependence becomes:

$$l(F) = z_0\left[\left[1 + 2\left(\frac{F}{F_\infty}\right)\left(\frac{r_0}{z_0}\right)\right]^{1/2} - 1\right] \qquad (14) \qquad \text{in which appears the ratio } r_0/z_0 = \alpha_0 \text{ already seen}$$

before and which is the predetermined numerical aperture of the beam, and the ratio $F/F_\infty$, an important ratio that can be called overfluence since it must be as large as possible to provide deep hole. Experimental length versus fluence is successfully simulated by equation (14) with adjusting of parameter $F_\infty$ for the given polymers (Table.2) up to obtaining a good fit.

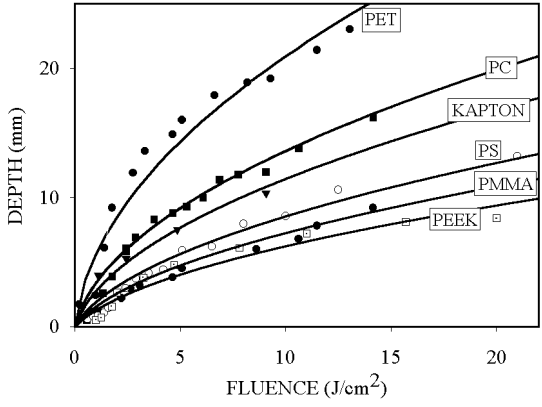

**Fig.10** Hole length at stationary profile versus fluence. Points are experimental measurements and solid line are obtained with model equation (14) fitted by adjusting $F_\infty$

**Table 2** Polymer characteristic threshold $F_\infty$ of ablation extinction at the end of drilling provided by the fit with model (14).

| Material | $F_\infty$ (mJ/cm²) |
|----------|---------------------|
| PET      | 1                   |
| PC       | 2.2                 |
| PEEK     | 8.5                 |
| PI       | 3                   |
| PS       | 5                   |
| PMMA     | 6.6                 |

In (14) we see dimensionless overfluence $F/F_\infty$ plays a particular role since a high value provides the good magnitude of drilling length and ensures the high aspect ratio obtained in this work (e.g. for PET it is $3 \times 10^4$ with $F_\infty$=1 mJ/cm² and laser fluence F=30 J/cm²). If the beam profile is taken as a gaussian profile [16]

$$F(\alpha) = F \exp(-2\alpha^2/w_\alpha^2) \qquad (15)$$

$$l(F) = z_0 \left[ \left[ 1 + 2 \left( \frac{F}{F_\infty} \right) \left( \frac{\sqrt{\pi} w_\alpha}{\sqrt{2}} \right) \right]^{1/2} - 1 \right] \qquad (16).$$

the length versus fluence dependence is only slightly modified:

$w_\alpha$ is the parameter in the gaussian formula.

### 3.4 Hole profile and beam transmission into the hole

With the help of equation (11), the calculated hole profile can be plotted as in Fig.11 for a beam having a flat-top profile with different values of fluence. Fluence can be expressed in unit of $F_{par} = F_\infty z_0/r_0$, which is a particular value of fluence at $r = r_0$, providing parallel side walls at the keyhole entrance (Fig.11 curve (2)). Calculated keyhole profiles of Fig.11 can be not only convergent but also divergent at the entrance and convergent in a deeper part (profiles 3-5 for example).

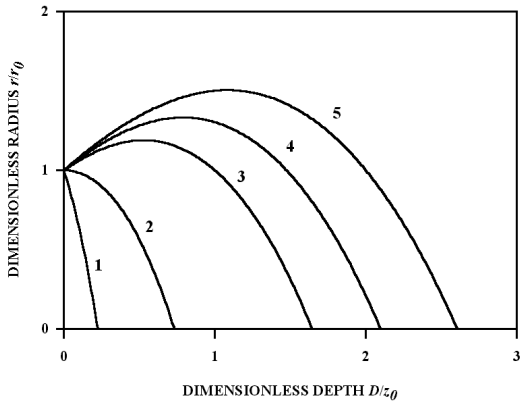

**FIG. 11.** Calculated stationary profiles for a number of laser fluences for top-hat distribution of incident beam, $r_0$ is beam radius at the entrance plane of the sample, $z_0$ is the distance between the focal point of the beam and the position of the material front surface, $D = z - z_0$. $F/F_{par} = :$ (1) 0.25; (2) 1; (3) 3; (4) 4.3; (5) 6. $F_{par} = F_\infty z_0/r_0$ is a threshold fluence at the spot border, $r = r_0$, providing parallel side walls at the keyhole entrance.

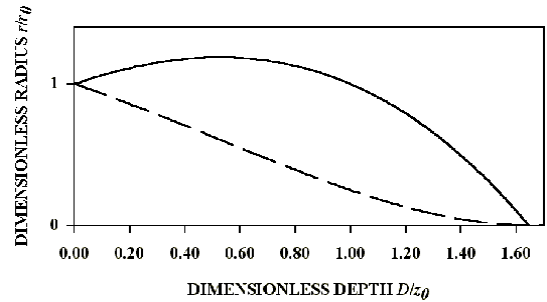

**FIG. 12.** Via hole transmission (dashed) vs. dimensionless hole depth for top-hat beam for $F = 3F_{par}$. Here $r_0$ is the beam radius at the entrance plane of the sample, $z_0$ is the distance between the beam waist and the position of the material front surface, $D = z - z_0$. The corresponding keyhole profiles are shown with solid curves.

In the multiple-scattering model all rays propagating from the point source to the side walls are totally absorbed, as we have $A_{eff}=1$. Therefore the only contribution to the transmission is given by the energy of the rays directly reaching the via hole exit from the point source O. For top-hat beam the transmitted energy $E(z)$ is then simply directly proportional to the solid angle at which the exit diameter is seen from the point source and the following transmission can be obtained:

$$T(z) = \frac{E(z)}{E(z_0)} = \frac{\pi r(z)^2 z_0^2}{\pi r_0^2 z^2} \qquad (17).$$

From Fig.12 we see that the calculated transmission curve is nearly a linear function of depth and does reproduce well the experimental measurements [16].

## 4. MECHAMISMS AND PERSPECTIVES

The present model is a good theoretical support for explaining some features of the experimental results, in particular for the prediction of the geometrical dimensions of the laser drilled hole when it has reached its final stationary profile. It gives a reasonable answer to the question which initiated these experiments, i.e. how deep a given laser beam can drill ? It defines a contour surface $r(\alpha)$, $z(\alpha)$ where $F_a(\alpha)=F_\infty$ and inside which the polymer is removed by ablation (ablative volume). It is important to note that it is independent of the pulsed character of the laser radiation. It is equally applicable to cw laser drilling provided the mechanism of material removal involves only a solid-to-gas transition in the absorbing volume when fluence is larger than the minimum for material removal $F_\infty$. But it is not intended to explain how the drilling performs in the intermediate steps. In the present pulsed laser drilling the process is repetitive with a period consisting of the following sequence of events: absorption, excitation, ablation, expansion and cooling. At the frequency of our laser (200 Hz) it is considered that complete cooling takes place between two pulses. It means that temperature goes back to initial one because heat diffuses fast enough (the surface cooling time is of the order of 1-10 ms) around the hole in a volume strictly limited to what can be reached by the energy content of a single pulse, that is to say only ~1-10² μm in most cases. The absorption step is then in general done on a new cold surface for each pulse but because of transmitted energy decreases with depth the end of drilling is reached when on each point of the keyhole surface $F(r,z) = F_\infty$. After some time of drilling, when ablation is deeper in the material, redeposition of the ablated products on the side walls comes into play but to a more or less extend depending on the nature of the polymer and its ablation mechanisms. For instance PMMA is typically ablating by a thermal mechanism with the formation of a substantial amount of liquid which can be expulsed under the form of droplets or redeposited on the walls. Redeposition and "re-ablation" is one of the two main reasons which reduce the drilling rate in Fig.1A. Another reason is the increased plume shielding due to a more 1D expansion for deep ablation rather than the 3D for surface ablation. Furthermore at this level of fluence this strongly absorbing and expanding plume, probably a transient plasma, plays also the role of storage of the incoming energy and can release it to the material surface during expansion. After each pulse the plasma energy transfer to the surface can have a cleaning or ablative effect on the redeposited products. It is also likely that if a liquid film is formed on the inner wall during drilling it is accelerated toward the exit of the drilled hole by momentum transfered from the outcoming gas plume. This may also contribute to an increase of hole diameter. The nature of the laser/matter interaction is mainly laser-surface but can be laser-plasma and plasma-surface with the later one being more delocalised over the entire surface of the hole and mainly over the side walls. Up to now little is known on the intimate mechanisms of material transport to the outside when the ablation front is deep in the material.

A good control on the laser beam leading to deep ablation or machining is suitable in many laser machining approaches. An experimental set-up identical to ours can be used for precise or complicated cutting materials over a large thickness just by a computer controlled moving of the target relative to the beam : see example Fig.13 in reference [12], helical drilling in reference [5]. Extensive microdrilling of multi-hole arrays is also needed for some applications. Higher frequency lasers [23] (up to ~5-50 kHz, 10 ps-30 ns, $I$ = 10-20 MW for instance) are becoming available and might be of interest if one looks for higher processing speed. The short excimer laser wavelengths (193 and 248 nm) are preferable in most cases, but 4ω and 3ω harmonics of the Nd/YAG laser are interesting sources with the highest frequencies. Such higher frequency laser may in principle offer a larger speed of processing. Nevertheless the new problem of target temperature increase may rise with increasing repetition rate of the laser pulse [23].

Other materials like metals, ceramics and glasses can also be treated by the KrF laser beam in a similar way [24]. However due to larger thresholds of ablation ($F_t$ and $F_\infty$) only much lower aspect ratio can be reached: 10-50. Furthermore the interest for using an ultraviolet laser is less obvious in this case. Many other sources of visible wavelength are being used for these particular materials like high power copper vapour lasers [25]. Also interesting work is done with femtosecond lasers since virtually liquid-free ablation in metal [26], silica [27], quartz [28] and polymers [29] drilling is reported. Ultraviolet femtosecond laser microdrilling of polymers is of basic interest and remains a perspective for the future.

## 5. CONCLUSIONS

We have presented recent experimental and theoretical work on high aspect ratio microdrilling of polymers with the pulsed KrF laser radiation. Good laser beams are obtained by condensing the laser on a circular molybdenum mask which in turn is imaged with a short focal projection lens onto the surface of the target. The best aspect ratio ∼ 600 is obtained in PET for a 15-20 µm hole diameter and 18 mm length. 6 different polymers have been studied. Hole length versus fluence dependence is predicted successfully with our model, which also provides a new threshold fluence $F_\infty$ minimum of energy density required for material removal. An important parameter for deep drilling is the overfluence ratio $F_0/F_\infty$ which determines the depth at which the drilling profile becomes stationary. This understanding will be helpful for many applications and future research.

## 6. ACKNOWLEDGEMENTS

The research is partly funded by Région Aquitaine and FEDER program. Authors thank CNRS (Centre National de Recherche Scientifique) of France for a research position for VNT, who is on leave from the General Physics Institute, Moscow.